\documentclass[aps,pre,twocolumn,float]{revtex4}
\usepackage{amsmath,bm,epsfig}
\usepackage{multirow}
\let \nn  \nonumber

\let\*\cdot
\def\<{\left\langle} \def\>{\right\rangle} \def\({\left(} \def\){\right)}
 \let\~\widetilde \let\^\widehat 

\def\be{\begin{equation}}\def\ee{\end{equation}}
\def\bea{\begin{eqnarray}}\def\eea{\end{eqnarray}}
\def\bse{\begin{subequations}}\def\ese{\end{subequations}}
\newcommand{\BE}[1]{\begin{equation}\label{#1}}
\newcommand{\BEA}[1]{\begin{eqnarray}\label{#1}}
\newcommand{\BSE}[1]{\begin{subequations}\label{#1}}

\let \nn  \nonumber
\usepackage{pstricks}
\usepackage{pst-node}
\usepackage[ansinew]{inputenc}
\usepackage{amssymb,amsmath}

\def\BSE{\begin{subequations}}\def\ESE{\end{subequations}}
\let \= \equiv

\def\o{\omega}
\def\wt{\widetilde}

\def\be{\begin{equation}}       \def\ba{\begin{array}}

\def\ee{\end{equation}}         \def\ea{\end{array}}

\def\bea {\begin{eqnarray}}      \def\eea {\end{eqnarray}}

\def\bean{\begin{eqnarray*}}    \def\eean{\end{eqnarray*}}

\def\eps{\varepsilon}           

\def\const {\mathop{\rm const}\nolimits}

\def\RA {\ \Rightarrow\ }

\def\<{\langle} \def\({\left(}  \def\>{\rangle} \def\){\right)}

\newtheorem{exi}{Example}

\begin{document}

\title{Energy spectra of \textbf{2D} gravity and capillary waves with narrow frequency band excitation}
\author{Elena Kartashova$^{\dag},$}
 \email{Elena.Kartaschova@jku.at}
  \affiliation{$^{\dag}$ Institute for Analysis, J. Kepler University, Linz, A-4040 Austria, EU}
  \affiliation{$^*$ Kavli Institute for Theoretical Physics, University of California, Santa Barbara, CA 93106, USA}
   \begin{abstract}
In this Letter we present a new method, called chain equation method (CEM), for computing  a cascade of distinct modes in a two-dimensional weakly nonlinear wave system generated by narrow frequency band excitation.  The CEM is a means for computing the quantized energy spectrum as an explicit function of  frequency $\o_0$ and stationary amplitude $A_0$ of excitation. The physical mechanism behind the generation of the quantized cascade is  modulation instability. The CEM can be used in numerous \textbf{2D} weakly nonlinear wave  systems with narrow frequency band excitation appearing in hydrodynamics, nonlinear optics, electrodynamics, convection theory  etc. In this Letter the CEM is demonstrated with examples of gravity and capillary waves with dispersion functions $\o(k) \sim k^{1/2}$ and $\o(k) \sim k^{3/2}$ respectively, and for two different levels of nonlinearity $\eps=A_0k_0$:  small ($\eps\sim 0.1$ to 0.25) and moderate ($\eps\sim 0.25$ to 0.4).
\end{abstract}

{PACS: 05.60.-k, 47.35.Bb, 47.35.Pq}


\maketitle

\textbf{1. Introduction.}

 The theory of weakly nonlinear wave interactions, also called wave turbulence theory, is based on the assumption that only resonant  or close to resonant (in some well defined sense) wave interactions have to be taken into account. Resonance conditions are written in the form
\bea
\o_1+\o_2=\o_3 +\Omega, \quad \vec{k}_1+\vec{k}_2=\vec{k}_3 + \Theta,  \label{3w-RC}\\
\o_1+\o_2=\o_3+\o_4+\Omega, \quad \vec{k}_1+\vec{k}_2=\vec{k}_3+\vec{k}_4 + \Theta,\label{4w-RC}\\
\quad 0 \le \Omega \ll 1, \quad \Theta \ge 0 \label{Omega-res}
\eea
for 3- and 4-wave systems correspondingly. Here $\vec{k}$ is wave vector, $\o=\o(\vec{k})$ is dispersion function (notation $\o_j=\o(\vec{k}_j)$ is used for brevity) while $\Omega$ and $\Theta$ is frequency and wave vector mismatch correspondingly. In this Letter we study only \textbf{2D} wave systems. Various scenarios of  energy transport in a \textbf{2D} wave system are known depending of the values of $\Omega$ and $\Theta$; some of them are shown in Table \ref{t:1} (detailed description of this classification will be given in a forthcoming paper).
\begin{table}[thc]
\begin{tabular}{|c|c|c|c|c|}
\hline
 &\multicolumn{1}{c|}{ \textbf{1}:  }&\multicolumn{1}{c|}{\textbf{2}: }\\
 &\multicolumn{1}{c|}{Narrow frequency }&\multicolumn{1}{c|}{Wide frequency }\\
 &\multicolumn{1}{c|}{band excitation}&\multicolumn{1}{c|}{range excitation}\\
\cline{2-3}
                      &  $0 \le \Omega <\delta_0 \ll 1$   &  $0 <\delta_0 < \Omega  \ll 1$  \\
\hline\hline
 \textbf{A}:      & set of  dynamical   &  \\
  $\Theta=0$     & systems, \cite{CUP} &  \\
\hline
 \textbf{B}:   & chain equation    &  wave kinetic  \\
  $\Theta >0$   & (\textbf{CEM})  & equation, \cite{ZLF92-Naz11}\\
\hline
\end{tabular}
\caption{Scenarios of energy transport in 3- and 4-wave \textbf{2D}-dimensional systems and corresponding references in literature, if available; otherwise the cell is left empty \label{t:1}}
\end{table}

\emph{\textbf{For wave systems with narrow frequency band excitation}} a theory exists for the resonance conditions (\ref{3w-RC})-(\ref{Omega-res}) with zero or small enough frequency mismatch  $0 \le \Omega < \delta_0 \ll 1$ and $\Theta=0$
 meaning that wave phases are coherent. These are the so-called exact and quasi resonances shown in the  Table \ref{t:1}, \textbf{A1}.

Evolution of the system over time  is described by a set of independent dynamical systems, one for each resonance cluster. In a 3-wave system the simplest clusters, which are called primary clusters,  are described by  dynamical systems of the form
\be  \label{3w-DS}
\dot{B}_1=  V^3_{12} B_2^*B_3,\,
\dot{B}_2=  V^3_{12} B_1^* B_3, \, \dot{B}_3=  - V^3_{12} B_1 B_2
\ee
with $B_j,\, j=1,2,3,$ denoting the wave amplitudes in canonical variables  and $V^3_{12}$ the coupling coefficient; differentiation  in (\ref{3w-DS})  is taken over the "slow" time $\tau=\eps t$ where $0<\eps \ll 1$ is a parameter of nonlinearity and time $t$ corresponds to the linear regime.

 All other clusters, called composite clusters, are  combinations of primary clusters having wave vectors in common, and their dynamical systems can be written out explicitly.
This also applies
to  4-wave systems, however the form of the primary  clusters is more complicated, and differentiation in the corresponding dynamical system is taken over the slow time $\tau=\eps^2 t$, \cite{CUP}.

In a 3-wave system excitation of the $\o_3$-mode (the high frequency mode)  yields  periodic energy exchange among the modes of a primary cluster; in a composite cluster also chaotic energy exchange is possible (Table \ref{t:1},  \textbf{A1}). Modes with smaller frequencies, $\o_1$- and $\o_2$-modes are neutrally stable; when excited they do not change their energy at the time scale $\tau=\eps t$.

In a 4-wave system  with
one-mode excitation resonance is only possible for the degenerate case with three different frequencies, \cite{Has67}. These quartets are called Phillips quartets and in fact behave like  a 3-wave system but with a different dynamical system
 (Table \ref{t:1}, \textbf{A1}).

\emph{\textbf{For wave  systems with wide range frequency excitation}} a theory exists only under the condition that
 frequency mismatch  $\Omega $ in (\ref{3w-RC})-(\ref{Omega-res}) is big enough to exclude exact and quasi resonances, in order to avoid the small divisor problem; wave phases are incoherent, Table \ref{t:1}, \textbf{B2}. Energy transport is governed by the wave kinetic equation derived from the statistical description of a wave system.  This approach   originates  in the pioneering paper of Hasselmann \cite{has1962}.
The 3-wave kinetic equation reads
\bea
\frac{\bf d }{{\bf d } t}{B}^2_k=
\int |V^k_{12}|^2
\delta(\o_k-\o_1-\o_2)\delta(\vec{k}-\vec{k}_1-\vec{k}_2) \nn \\
\cdot (B_1B_2-B_1^{*}B_k-B_2^{*}B_k)
{\bf d}\vec{k}_1 {\bf d}\vec{k}_2 \label{3w-KE}
\eea
(it is written for average quantities $<B_k B_k*>$ and  resonance conditions of the form (\ref{3w-RC}) where $\vec{k}$ takes all possible values $\vec{k}_3$); the 4-wave kinetic equation has a similar form (not shown here for place). The stationary solution of the corresponding wave kinetic equation
is the continuous energy spectrum, \cite{ZLF92-Naz11}.

Quantized spectra observed in the numerous laboratory experiments with narrow frequency band excitation \cite{BF-Taiw,denis,NR11,XiSP10}, etc.  have not yet been described theoretically.

In this Letter we present a method based on a chain equation (chain equation method, CEM)  to
compute the form of the quantized energy spectrum  as an explicit function of frequency $\o_0$ and stationary amplitude $A_0$ of excitation (Table \ref{t:1}, \textbf{B1}). The CEM is demonstrated with examples of  capillary and gravity waves  (usually regarded as pure 3- and  4-wave system respectively)  in order to show that the underlying mechanism of quantized cascade generation in both types of system is the same, namely modulation instability which is a 4-wave mechanism.\\

\textbf{2. Outline of the CEM}

A  model describing the generation of distinct modes (quantized cascade) in a gravity water wave system with narrow frequency band excitation has  been presented in
\cite{KSh11}. In \cite{KSh11} several fundamental aspects of Benjamin-Feir instability are explained qualitatively
(e.g. dependence of cascade direction on  excitation parameters and asymmetry of amplitudes of side-bands ) under the assumption that the frequency shift $(\Delta \o)_n$ between neighboring cascade steps is constant with respect to $n$. 
In the present Letter we relax this assumption. This allows us to compute the quantized energy spectrum.

Benjamin-Feir or modulation instability (known in various physical areas under different names such as parametric instability in classical mechanics, Suhl instability  of spin waves, Oraevsky-Sagdeev  decay instability of plasma waves,  etc. \cite{ZO08}) is the decay of a carrier wave  $\o_0$ into two neighboring side band modes $\o_1, \, \o_2$ described by:
\bea
\o_1 + \o_2 = 2\o_0, \quad
\vec{k}_1+\vec{k}_2=2\vec{k}_0+ \Theta, \label{ModInst}\\
 \o_1=\o_0 + \Delta \o, \, \o_2=\o_0 - \Delta \o, \,  0<\Delta \o \ll 1. \label{Delta-Omega} \eea
 The growth of modulation instability is characterized by the instability increment.
 In the seminal work \cite{BF67}  Benjamin and Feir have shown (for gravity waves with small initial steepness, $\eps\sim 0.1$ to 0.2) that
a wave train with initial real amplitude $A$, wavenumber $k=|\vec{k}|$, and frequency $\o$  is unstable to perturbations with a small frequency mismatch $\Delta \o$, when the following condition is satisfied:
$
0 \le {\Delta \o}/{A k\o} \le \sqrt{2}\nn
$
and the increment has its  maximum for
\be \label{BFI-incr}
\Delta \o / A k \o =1.
\ee
 Similar conditions for other wave systems and moderate nonlinearity can be found e.g.  in \cite{DY79,Hog85}.

\noindent Notice that $\Delta \o=\o_1-\o_0 $ or $\Delta \o=\o_0-\o_2$ for direct and inverse energy cascades correspondingly which means that there are
two different expressions:
one for direct cascade and one for inverse cascade for computing the frequency at which the maximum of the instability increment is achieved.
\noindent The main steps of the chain equation method (CEM) are performed as follows, separately for direct and inverse cascade (depending on the excitation parameters, in any of those passes a cascade may be generated):

\noindent \emph{Step 1}. \emph{Compute the frequencies of the cascading modes:}

\noindent  At the first cascade step $n=1$, according to (\ref{Delta-Omega}) from the excitation frequency $\o_0$  the frequency
of the cascading mode $\o_1$ is computed such that its increment of instability is maximal. In each subsequent step the frequency of the previous cascading mode $\o_{n-1}$ is regarded as the excitation frequency and a new cascading mode with frequency $\o_{n}$ is generated.

\noindent \emph{Step 2}. \emph{Deduce the chain equation:}

\noindent The chain equation  is a recursive relation between subsequent cascading modes involving frequencies $\o_n, \, \o_{n+1}$ and amplitudes $A_n, \, A_{n+1}$. It is derived using the explicit form of the condition of maximal instability increment, assuming that the fraction $p$ of energy transported from one cascading mode to the next one depends only on the excitation parameters and not on the cascade step.

\noindent \emph{Step 3}. \emph{Compute the amplitudes of the cascading modes:}

\noindent Taylor expansion of the chain equation is taken; its first two terms provide an ordinary differential equation whose solution is a quantized energy spectrum.\\

\noindent The method is described in detail in the next section.\\

\textbf{3. The CEM.}

\textbf{\emph{3.1. Assumptions.}} We assume that

\textbf{(*)} $p_n=p=\const$, i.e. \emph{the cascade intensity} $p$ for a given set of the excitation parameters is constant.

\textbf{(**)} at each cascade step $n$, the cascading mode with frequency $\o_{n}$ has maximal increment of instability.

 Assumption (*) is confirmed by experimental data, e.g.  for capillary water waves \cite{XiSP10},  vibrating elastic plate \cite{MorPers}, gravity water waves \cite{KSh11-2}. Assumption (**) can be regarded as a reformulation of
 the hypothesis of Phillips  that "the spectral density  is saturated at a level determined by wave breaking" (§2.2.6, \cite{Jans04}).

\textbf{\emph{3.2. Step 1: Computing frequencies}}

\noindent Let us regard a cascading chain of the form
\bea \label{general}
\begin{cases}
\o_{1,1}+\o_{2,1}=2\o_{0},    &E_1=p_1E_0, \, , \\
\o_{2,1}+\o_{2,2}=2\o_{1,1},  &E_2=p_2E_1, \\
....  \\
\o_{n,1}+ \o_{n,2}=2\o_{n-1,1},   &E_n =p_n E_{n-1}
\end{cases}
\eea
where (\ref{Delta-Omega}) is satisfied at each cascade step $n$, with $(\Delta \o)_n$ depending on $n$. For brevity, frequencies of the cascading modes $\o_{1,j}$ are further  notated as $\o_{j}$. Here
$\o_0$ is the excitation  frequency, $E_n$ is the energy  at the $n$-th step of the cascade and $p_n, \, 0< p_n < 1,$ denotes the fraction of the energy $E_{n-1}$ transported  from the cascading mode with amplitude $A_{n-1}=A(\o_{n-1})$ to the cascading mode with amplitude $A_n=A(\o_{n})$.  Accordingly
\be
\o_0 <\o_{1} < \o_{2} < \o_{3}< ... <\o_{n}< ...
\ee
\noindent for direct cascade and
\be
\o_0 > \o_{1} > \o_{2} > \o_{3}> ... >\o_{n}> ...
\ee
for inverse cascade.

\textbf{\emph{3.3. Step 2: Deducing chain equation.}}

\noindent Construction of a chain equation is demonstrated below for the example of gravity water waves where  conditions providing maximum of the instability increment hold as  given by (\ref{BFI-incr}).
The example was chosen for its simple form. For demonstrating that our method is quite general, in Sec.4, 5 more complicated examples are given.

 It follows from (*) that
\be \label{p}
E_n =p E_n \RA A_{n+1}=\sqrt{p} A_n.
\ee
Condition (\ref{BFI-incr}) is now rewritten as  $ (\Delta \o)_n/\o_n A_n k_n= 1 $ and yields $ (\Delta \o)_n$.

From assumption (**) we obtain the frequency of the next cascading  mode
 as
\be \label{inc-n1-dir}
 \o_{n+1}=\o_n +  \o_n A_n k_n
\ee
 for direct cascade and
\be \label{inc-n1-inv}
 \o_{n+1}=\o_n -  \o_n A_n k_n
\ee
 for inverse cascade.
In this sense, we may combine  (\ref{inc-n1-dir}) and (\ref{inc-n1-inv}) into
\be \label{inc-n1}
 \o_{n+1}=\o_n \pm \o_n A_n k_n.
\ee
Thus, as all calculations for direct and inverse cascades are identical save the sign of the second term,
in all equations below we write "$\pm$" at the corresponding place, understanding that "+" should be taken for direct cascade and "-" for inverse cascade. Combination of (\ref{p}) and (\ref{inc-n1}) yields $\sqrt{p} A(\o_n)= A(\o_{n+1})$, i.e.
\be \label{1}
\sqrt{p} A_n=  A(\o_n \pm \o_n A_n k_n)
\ee
which is further on called \emph{the chain equation}.

\textbf{\emph{3.4. Step 3: Computing amplitudes.}}

\noindent Let us take the Taylor expansion of  RHS of the chain equation:
\bea \label{Taylor}
\sqrt{p} A_n= A(\o_n \pm \o_n A_n k_n)= \sum_{s=0} ^ {\infty } \frac {A_n^{(s)}}{s!} \, (\pm\o_n A_n k_n)^{s} \nonumber \\
 = A_n \pm  A_n^{'}\o_n A_n k_n+ \frac 12 A_n^{''}(\pm \o_n A_n k_n)^2+...
\eea
 with differentiation over $\o_n$ and $k_n=k(\o_n)$ as defined by the dispersion relation.

Taking two first RHS terms from  (\ref{Taylor}) we obtain
\bea
\sqrt{p} A_n \approx A_n \pm  A_n^{'}\o_n A_n k_n
\RA A_n^{'} = \pm \frac{\sqrt{p}-1}{\o_n k_n } \RA  \label{2terms}  \\
A(\o_n) = \pm (\sqrt{p}-1) \int \frac{d \o_n}{\o_n k_n}+C(\o_0, A_0, p) \qquad \label{A-gen}
\eea
 where $\o_0, A_0$ are excitation parameters and $p=p(\o_0, A_0)$. Accordingly, energy $E(\o_n) \sim A^2(\o_n).$\\

\textbf{4. Gravity water waves, $\o^2 \sim k$.}

\textbf{\emph{4.1. Gravity water waves with small nonlinearity,}} $\eps\sim 0.1$ to 0.25.

It follows from (\ref{A-gen}) that for \emph{direct cascade}
\bea
 A(\o_n) = (\sqrt{p}-1) \int \frac{ d \o_n}{\o_n^3} = \frac{(1-\sqrt{p})}{2} \o_n^{-2} + C^{(Dir)},\\
C^{(Dir)} = A_0 - \frac{(1-\sqrt{p})}{2} \o_0^{-2},
\eea
 which yields an energy spectrum of the form
\be \label{energy-grav-Direct}
E(\o_n)^{(Dir)} \sim \Big[ \frac{(1-\sqrt{p})}{2} \o_n^{-2}+C^{(Dir)} \Big]^2.
\ee
Computations of energy spectra for \emph{inverse cascade} are analogous and yield
\bea
E(\o_n)^{(Inv)} \sim \Big[ -\frac{(1-\sqrt{p})}{2} \o_n^{-2} +C^{(Inv)} \Big]^2, \label{energy-grav-Inverse}\\
C^{(Inv)} = A_0 + \frac{(1-\sqrt{p})}{2} \o_0^{-2}. \label{grav-inv-const}
\eea

In particular,  for specially chosen excitation parameters
$
\o_0^{-2}(1-\sqrt{p})=2A_0
$
the direct cascade has an infinite number of steps, with quantized energy spectrum $E_{\infty}(\o)^{(\mathbf{Dir})}$ having the form
\be
E_{\infty}(\o_n)^{(Dir)} \sim   \o_n^{-4}. \label{casc-infty}
\ee
In this case  spectral density $S(E(\o))$ showing how fast the spectrum falls can be computed as follows:
\bea S(E(\o)) &=& \lim_{\o_{n+1}-\o_{n} \rightarrow 0} \frac{E(\o_{n+1})-E(\o_{n})}{\o_{n+1}-\o_{n}}\nn \\
&=&\frac{d E(\o)}{d \o} \sim \o^{-5}.\eea
 This is in accordance with the JONSWAP spectrum which is an empirical relationship based on  experimental data.

Conditions for  direct ($\o_{n+1}>\o_n$) and inverse ($\o_{n+1}<\o_n$) cascade to occur can be studied in a similar way as in \cite{KSh11}.
Cumbersome formulae for computing $p$ as function of initial parameters are not shown for space; they shall be presented in \cite{KSh11-2}.
 Notice that cascade intensity $p$ can be easily determined from experimental data.

\textbf{\emph{4.2. Gravity water waves with moderate nonlinearity,}} $\eps\sim 0.25$ to 0.4.

Maximal instability increment is achieved in the case of moderate steepness ($\varepsilon \sim 0.25$ to 0.4) if, \cite{DY79},
\be \label{Dys}
\Delta \o/\Big(\o  A k - \frac{3}{2}\o^2  A^2 k^2\Big)=1.
\ee
In this case assumptions (*),(**) yield
\bea
&&\begin{cases}\nn
E_n =p E_n \RA A_{n+1}=\sqrt{p} A_n,\\
|(\Delta \o)_n|/\Big(\o_n A_n k_n - \frac{3}{2}(\o_n A_n k_n)^2\Big)=1.
\end{cases} \RA  \\
&&\sqrt{p}-1 \approx \pm  A_n^{'}\o_n^3 \mp \frac{3}{2}A_n^{'}A_n \o_n^6,\label{ODE-Dys}
\eea
where again only two terms of the corresponding Taylor expansion are taken into account (upper signs for direct cascade; lower signs for inverse cascade).

The general solution of (\ref{ODE-Dys}) has the form  $A_n=\wt{A}_n+\wt{\wt{A}}_n$ where $\wt{A}_n$ is the general solution of the homogeneous part of (\ref{ODE-Dys}) and $\wt{A}_n$ is a particular solution of (\ref{ODE-Dys}) (taken below at $\o_n=1$). Accordingly
\be
\pm  \wt{A}_n^{'}\o_n^3 \mp \frac{3}{2}\wt{A}_n^{'}\wt{A}_n \o_n^6=0 \RA \wt{A}_n= \frac{3}{2}\o_n^{-3}
\ee
 while equation
\be
\pm \wt{\wt{A}}_n \mp \frac{3}{2}\wt{\wt{A}}_n^{'}\wt{\wt{A}}_n=\sqrt{p}-1
\ee
 has two solutions
\be
\wt{\wt{A}}_n = \frac{1}{3} \left(2 \pm \sqrt{2} \sqrt{2-6 (\sqrt{p}-1)\o_n+3 \mathcal{C}^{(Dir)}}\right),\label{An-dir}
\ee
 for direct cascade and two solutions
\be
\wt{\wt{A}}_n = \frac{1}{3} \left(2 \pm \sqrt{2} \sqrt{2+6 (\sqrt{p}-1)\o_n+3 \mathcal{C}^{(Inv)}}\right),\label{An-inv}
\ee
for inverse cascade.

To determine the sign in (\ref{An-dir}),(\ref{An-inv}) we assume without loss of generality that $\o_0=1,$ implying $\o_n>1$ for direct cascade and $\o_n<1$ for inverse cascade. As the energy of the modes decreases with growth of $n$ we get finally
\bea
\mathcal{E}(\o_n)^{(Dir)} \sim \Big[\frac{3}{2}\o_n^{-3}  \nn \\
+\frac{1}{3} \left(2 - \sqrt{2} \sqrt{2 - 6 (\sqrt{p}-1)\o_n+3 \mathcal{C}^{(Dir)}}\right)\Big]^2, \label{energy-grav-Direct-Dys}\\
\mathcal{E}(\o_n)^{(Inv)} \sim \Big[\frac{3}{2}\o_n^{-3} \nn \\
 +\frac{1}{3} \left(2 + \sqrt{2} \sqrt{2 + 6 (\sqrt{p}-1)\o_n+3 \mathcal{C}^{(Inv)}}\right)\Big]^2.\label{energy-grav-Inverse-Dys}
\eea
\begin{figure}
\begin{center}\vskip -0.1cm
\includegraphics[width=7cm]{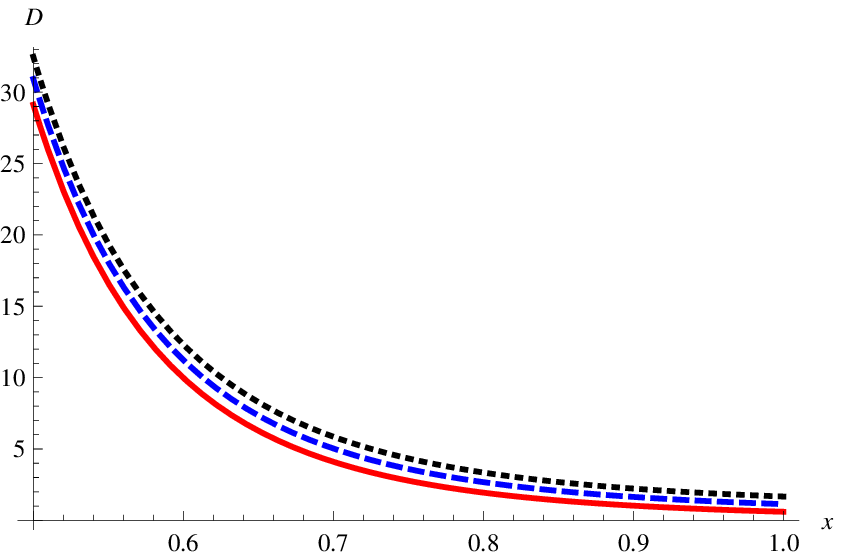}
\includegraphics[width=7cm]{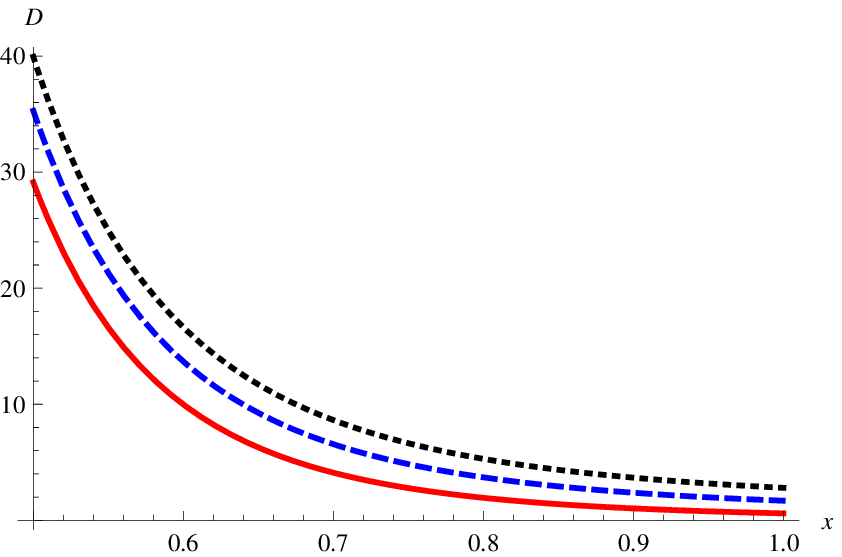}
\caption{\label{f:FunctionD} Color online. Plot of the function $D(x), \, 0.5 \le x \le 0.9$ for different values of parameters $p$ and $C$. \textbf{Upper panel:} $D(x)$ is shown for $1-\sqrt{p}=0.1; 0.5; 0.9$  (red bold, blue dashed and black dotted lines correspondingly); $C=0.1$. \textbf{Lower panel:} $D(x)$ is shown for $\sqrt{p}-1=0.1$, $C=0.1; 5; 10$  (red bold, blue dashed and black dotted lines correspondingly).}
\end{center}\vskip -0.7cm
\end{figure}
\begin{figure}
\begin{center}\vskip -0.1cm
\includegraphics[width=7cm]{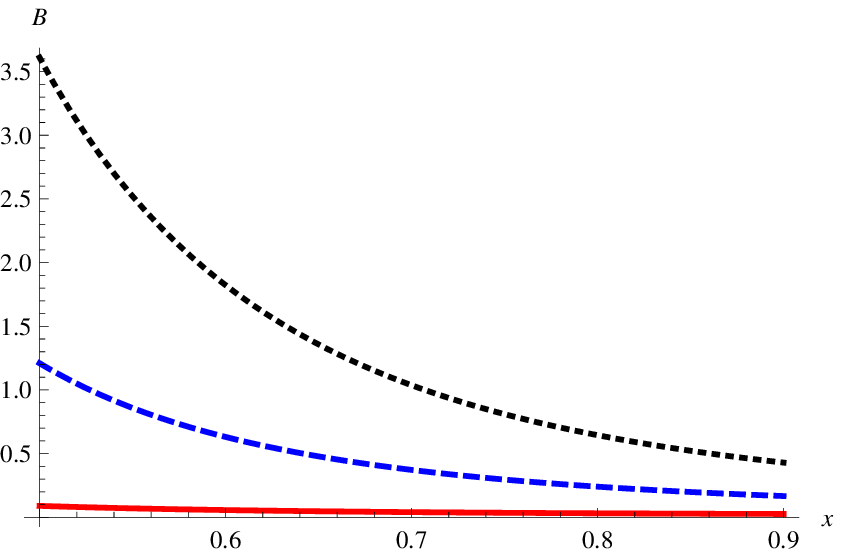}
\includegraphics[width=7cm]{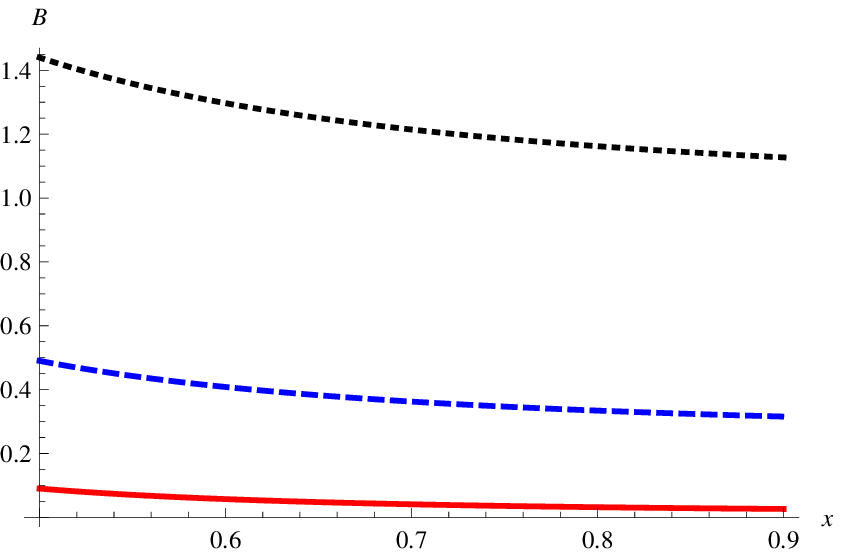}
\caption{\label{f:FunctionB} Color online. Plot of the function $B(x), \, 0.5 \le x \le 0.9$ for different values of parameters $p$ and $C$. \textbf{Upper panel:} parameters and color scheme as in Fig.\ref{f:FunctionD}, upper panel; \textbf{Lower panel:} $B(x)$ is shown for $\sqrt{p}-1=0.1$, $C=0.1; 0.5; 1$ (red bold, blue dashed and black dotted lines correspondingly).}
\end{center}\vskip -0.7cm
\end{figure}

We introduce the function
$D(x)=$
\be \label{D(x)}\Big[\frac{3}{2}x^{-3} +\frac{1}{3} \left(2 - \sqrt{2} \sqrt{2 - 6 (\sqrt{p}-1)x+3 C}\right)\Big]^2\ee
which describes the energy spectrum as given in (\ref{energy-grav-Direct-Dys}) and study its behavior for different values of parameters
 $p$ and $C$ (Fig.\ref{f:FunctionD}).

To compare energy spectra $\mathcal{E}_n(\o)$ and $E_n(\o)$  (for direct cascade) in Fig.\ref{f:FunctionB} we plot the function
\be \label{B(x)}B(x)=\Big[{(1-\sqrt{p})}x^{-2}/{2} +C \Big]^2.\ee
 In both Figs.\ref{f:FunctionD},\ref{f:FunctionB} in the upper panels $C$ is fixed,  while cascade intensity $p$ varies; in the lower panels cascade intensity $p$ is fixed and $C$ varies.

A simple observation can be made immediately: dependence of quantized energy spectra on  the excitation parameters is substantially smaller in a wave system with bigger nonlinearity. Indeed, for $C=0.1$ (Figs.\ref{f:FunctionD},\ref{f:FunctionB}; upper panels), case of moderate nonlinearity,
\be \Big(D(0.5)\Big |_{p=0.81}\Big)/\Big(D(0.5)\Big |_{p=0.01}\Big) \approx 0.9. \ee
This means a change of cascade intensity of almost of two orders of magnitude yields  $\sim 10\%$ change in the energy  at $\o = 0.5$. On the other hand, for the same
set of values $p,$
\be \Big(B(0.5)\Big |_{p=0.81}\Big)/\Big(B(0.5)\Big |_{p=0.01}\Big) \approx 0.03. \ee
This means  $B(x)$ changes by $300\%.$

A similar effect is observed if the cascade intensity is fixed, $p=0.81$, (Figs.\ref{f:FunctionD},\ref{f:FunctionB}; lower panels):
\bea \Big(D(0.5)\Big |_{C=0.1}\Big)/\Big(D(0.5)\Big |_{C=10}\Big) \approx 0.73,\\
\Big(B(0.5)\Big |_{C=0.1}\Big)/\Big(B(0.5)\Big |_{C=1}\Big) \approx 0.07. \eea
For the improvement of graphical presentation  function $B(x)$ is shown with a smaller value of $C$ than that used for function $D$
($B(0.5)\Big |_{C=10} \approx 104$, not shown in Fig. \ref{f:FunctionB}).

  From these figures preliminary conclusion can be made: dependence of the quantized energy spectrum on the parameters of excitation is lower for higher
  nonlinearity.\\

\textbf{5. Capillary waves,  $\o^2 \sim k^3$.}

Conditions for small and moderate nonlinearity read
\bea
(\Delta \o)_n /\Big(\frac{1}{24} \o_n A_n k_n \Big)=1, \label{cap-small}\\
(\Delta \o)_n/\Big(\frac{1}{24} \o_n A_n k_n+ \frac{3}{2}\o_n^2  A_n^2 k_n^2\Big)=1.
\eea
as is shown in \cite{Hog85}; all computations are completely similar to the examples presented in the previous section. Resulting formulae are shown below, for the case of
\emph{small nonlinearity}:
\bea
 \pm  (\sqrt{p} -1) \approx \frac{1}{24} A_n^{'}\o_n^{5/3}  \RA \nn \\
E(\o_n)^{(Dir)} \sim \Big[ \frac{(1-\sqrt{p})}{16} \o_n^{-2/3}+C^{(Dir)} \Big]^2,\label{cap-D-small_E}\\
E(\o_n)^{(Inv)} \sim \Big[ -\frac{(1-\sqrt{p})}{16} \o_n^{-2/3} +C^{(Inv)} \Big]^2,\\
\mbox{where}\qquad C^{(Dir)} = A_0 - \frac{(1-\sqrt{p})}{16} \o_0^{-2/3},  \\
C^{(Inv)} = A_0 + \frac{(1-\sqrt{p})}{16} \o_0^{-2/3}.
\eea
and for the case of \emph{moderate  nonlinearity}:
\bea
 \pm  (\sqrt{p} -1) \approx   \frac{1}{24}A_n^{'}\o_n^{5/3} + \frac{3}{2}A_n^{'}A_n\o_n^5, \RA \nn \\
\mathcal{E}(\o_n)^{(Dir)} \sim \Big[-\frac{1}{36}\o_n^{-5/2}  \nn \\
+\frac{1}{36} \left(-1 - \sqrt{1 - 1728 (\sqrt{p}-1)\o_n+72 \mathcal{C}^{(Dir)}}\right)\Big]^2, \\
 \mathcal{E}(\o_n)^{(Inv)} \sim \Big[-\frac{1}{36}\o_n^{-5/2}  \nn \\
+\frac{1}{36} \left(-1 + \sqrt{1 + 1728 (\sqrt{p}-1)\o_n+72 \mathcal{C}^{(Inv)}}\right)\Big]^2
\eea
where $\mathcal{C}^{(Dir)}$ and $\mathcal{C}^{(Inv)}$  are derived from the conditions
\bea
A_0=\frac{1}{36}\o_0^{-5/2}  \nn \\
-\frac{1}{36} \left(-1 - \sqrt{1 - 1728 (\sqrt{p}-1)\o_0+72 \mathcal{C}^{(Dir)}}\right), \\
A_0=\frac{1}{36}\o_0^{-5/2}  \nn \\
-\frac{1}{36} \left(-1 + \sqrt{1 + 1728 (\sqrt{p}-1)\o_0+72 \mathcal{C}^{(Inv)}}\right).\label{cap-I-mod-const}
\eea

\textbf{6. Conclusions.}
In this Letter the following results are presented:

\textbf{I.}~  The chain equation method (CEM) presented allows to compute quantized energy spectra  as a solution of an ordinary differential equation. Generation of cascading modes is only possible by modulation instability and only if exact and quasi-resonances do not occur. The dispersion function and the excitation parameters determine
the form of the differential equation and the resulting quantized  spectra.  Direct, inverse or both cascades are possible with a finite or infinite number of modes. All these scenarios have been found in experiment.

\textbf{II.}~In the examples given the CEM is applied for computing quantized energy spectra of gravity and capillary waves, and two different levels of nonlinearity. It is also shown that transition from quantized to continuous spectrum in the case of gravity water waves (for direct cascade) yields spectral density $\sim \o^{-5} $ which is in accordance with the notorious  JONSWAP spectrum empirically deduced from oceanic measurements.

The CEM has been applied to  more complicated forms of the dispersion function, e.g. for gravity-capillary waves with  $\o^2=g\, k + \sigma \, k^3$, which is not shown in the examples.

\textbf{III.}~Comparing the methods of chain equation and kinetic equation, one can see
from  Table \ref{t:1} that they  apply to  different sets of parameters and excitation form. Moreover, they make different predictions.
CEM predicts that  the form of the energy spectrum depends on the amplitude $A_0$ and  frequency $\o_0$ of excitation, whereas kinetic equation predicts that there is no such dependency. If one is unsure which method to apply,  it will be sufficient to measure the energy spectrum with two different sets of excitation parameters (e.g. with the same excitation frequency but different amplitude of excitation as it is done in \cite{XiSP10}).

\textbf{IV.}~Conditions of  quantized cascade termination can be derived from the chain equation. As they are not discussed in this Letter
  a short account is given below (detailed study of the direct cascade will be performed in \cite{KSh11-2}):

(a) In the case of direct cascade, transition to continuous spectrum  is possible with $n \rightarrow \infty$ (example is given by (\ref{casc-infty})) for certain values of excitation parameters. For other choices of excitation parameters, a
direct cascade terminates at some finite step $n$ due to one of the following reasons

- \emph{decrease of amplitudes}: amplitude of mode $A_n$ is so small that the wave system becomes linear which means no mode $A_{n+1}$ is produced; experimentally observed in \cite{BF-Taiw}.

- \emph{limiting steepness of total wave package}: with each additional cascading mode the resulting wave package becomes steeper until
the wave system becomes strongly nonlinear and wave breaking occurs;  experimentally observed in \cite{denis}.

- \emph{exact resonance among cascading modes}: if a mode $A_n$ generates an exact or quasi-resonance with some of the previously generated modes (Table \ref{t:1}, \textbf{A1}), (quasi) recurrent energy exchange among these modes will occur similar to the Fermi-Pasta-Ulam phenomenon and no new mode will be generated; experimentally observed in \cite{BF-Taiw}.

(b) An inverse cascade is always finite. It can be shown that under certain conditions, it terminates in a final mode with a frequency close to zero. Our  hypothesis is that this is the "zero-frequency sideband" observed in many experiments  \cite{XiSP10}. Moreover, it may be assumed that this mode interacts with previously generated modes thus yielding (at some new time scale) distributed initial state for kinetic regime to occur.

{\textbf{Acknowledgements.}} Author acknowledges  W. Farrell,
S. Lukaschuk,  W. Munk, M. Shats, I. Shugan and H. Tobisch  for valuable discussions and anonymous  Referees for useful remarks and recommendations. This research has been supported by the Austrian Science Foundation (FWF) under project P22943-N18 "Nonlinear resonances of water waves" and in part by the National Science Foundation, USA, under Grant No. NSF PHYS05-51164.

\end{document}